\documentclass[twocolumn,showpacs,aps,prl,superscriptaddress]{revtex4}

\usepackage[dvips]{graphicx}
\usepackage{latexsym}

\def\bslantfrac#1#2{{#1}\backslash\kern-0.1em{#2}}

\begin{document}

\newcommand{\Real}{\Re\text{e}}
\newcommand{\Imag}{\Im\text{m}}

%
%

\title{Deeply Virtual Compton Scattering off the Neutron}

\author{M.~Mazouz}
\affiliation{LPSC, Universit\'e Joseph Fourier, CNRS/IN2P3, INPG, F-38026 Grenoble, France}
\author{A.~Camsonne}
\affiliation{LPC Clermont-Ferrand, Universit\'e Blaise Pascal, CNRS/IN2P3, F-63177 Aubi\`ere, France}
\author{C.~Mu\~noz~Camacho}
\affiliation{CEA Saclay, DAPNIA/SPhN, F-91191 Gif-sur-Yvette, France}
\author{C.~Ferdi}
\affiliation{LPC Clermont-Ferrand, Universit\'e Blaise Pascal, CNRS/IN2P3, F-63177 Aubi\`ere, France}
\author{G.~Gavalian}
\affiliation{Old Dominion University, Norfolk, Virginia 23508, USA}
\author{E.~Kuchina}
\affiliation{Rutgers, The State University of New Jersey, Piscataway, New Jersey 08854, USA}
\author{M.~Amarian}
\affiliation{Old Dominion University, Norfolk, Virginia 23508, USA}
\author{K.~A.~Aniol}
\affiliation{California State University, Los Angeles, Los Angeles, California 90032, USA}
\author{M.~Beaumel}
\affiliation{CEA Saclay, DAPNIA/SPhN, F-91191 Gif-sur-Yvette, France}
\author{H.~Benaoum}
\affiliation{Syracuse University, Syracuse, New York 13244, USA}
\author{P.~Bertin}
\affiliation{LPC Clermont-Ferrand, Universit\'e Blaise Pascal, CNRS/IN2P3, F-63177 Aubi\`ere, France}
\affiliation{Thomas Jefferson National Accelerator Facility, Newport News, Virginia 23606, USA}
\author{M.~Brossard}
\affiliation{LPC Clermont-Ferrand, Universit\'e Blaise Pascal, CNRS/IN2P3, F-63177 Aubi\`ere, France}
\author{J.-P.~Chen}
\affiliation{Thomas Jefferson National Accelerator Facility, Newport News, Virginia 23606, USA}
\author{E.~Chudakov}
\affiliation{Thomas Jefferson National Accelerator Facility, Newport News, Virginia 23606, USA}
\author{B.~Craver}
\affiliation{University of Virginia, Charlottesville, Virginia 22904, USA}
\author{F.~Cusanno}
\affiliation{INFN/Sezione Sanit\`{a}, 00161 Roma, Italy}
\author{C.W.~de~Jager}
\affiliation{Thomas Jefferson National Accelerator Facility, Newport News, Virginia 23606, USA}
\author{A.~Deur}
\affiliation{Thomas Jefferson National Accelerator Facility, Newport News, Virginia 23606, USA}
\author{R.~Feuerbach}
\affiliation{Thomas Jefferson National Accelerator Facility, Newport News, Virginia 23606, USA}
\author{J.-M.~Fieschi}
\affiliation{LPC Clermont-Ferrand, Universit\'e Blaise Pascal, CNRS/IN2P3, F-63177 Aubi\`ere, France}
\author{S.~Frullani}
\affiliation{INFN/Sezione Sanit\`{a}, 00161 Roma, Italy}
\author{M.~Gar\c con}
\affiliation{CEA Saclay, DAPNIA/SPhN, F-91191 Gif-sur-Yvette, France}
\author{F.~Garibaldi}
\affiliation{INFN/Sezione Sanit\`{a}, 00161 Roma, Italy}
\author{O.~Gayou}
\affiliation{Massachusetts Institute of Technology,Cambridge, Massachusetts 02139, USA}
\author{R.~Gilman}
\affiliation{Rutgers, The State University of New Jersey, Piscataway, New Jersey 08854, USA}
\author{J.~Gomez}
\affiliation{Thomas Jefferson National Accelerator Facility, Newport News, Virginia 23606, USA}
\author{P.~Gueye}
\affiliation{Hampton University, Hampton, Virginia 23668, USA}
\author{P.A.M.~Guichon}
\affiliation{CEA Saclay, DAPNIA/SPhN, F-91191 Gif-sur-Yvette, France}
\author{B.~Guillon}
\affiliation{LPSC, Universit\'e Joseph Fourier, CNRS/IN2P3, INPG, F-38026 Grenoble, France}
\author{O.~Hansen}
\affiliation{Thomas Jefferson National Accelerator Facility, Newport News, Virginia 23606, USA}
\author{D.~Hayes}
\affiliation{Old Dominion University, Norfolk, Virginia 23508, USA}
\author{D.~Higinbotham}
\affiliation{Thomas Jefferson National Accelerator Facility, Newport News, Virginia 23606, USA}
\author{T.~Holmstrom}
\affiliation{College of William and Mary, Williamsburg, Virginia 23187, USA}
\author{C.E.~Hyde}
\affiliation{LPC Clermont-Ferrand, Universit\'e Blaise Pascal, CNRS/IN2P3, F-63177 Aubi\`ere, France}
\affiliation{Old Dominion University, Norfolk, Virginia 23508, USA}
\author{H.~Ibrahim}
\affiliation{Old Dominion University, Norfolk, Virginia 23508, USA}
\author{R.~Igarashi}
\affiliation{University of Saskatchewan, Saskatchewan, SK, Canada, S7N 5C6}
\author{X.~Jiang}
\affiliation{Rutgers, The State University of New Jersey, Piscataway, New Jersey 08854, USA}
\author{H.S.~Jo}
\affiliation{IPN Orsay, Universit\'e Paris Sud, CNRS/IN2P3, F-91406 Orsay, France}
\author{L.J.~Kaufman}
\affiliation{University of Massachusetts Amherst, Amherst, Massachusetts 01003, USA}
\author{A.~Kelleher}
\affiliation{College of William and Mary, Williamsburg, Virginia 23187, USA}
\author{A.~Kolarkar}
\affiliation{University of Kentucky, Lexington, Kentucky 40506, USA}
\author{G.~Kumbartzki}
\affiliation{Rutgers, The State University of New Jersey, Piscataway, New Jersey 08854, USA}
\author{G.~Laveissiere}
\affiliation{LPC Clermont-Ferrand, Universit\'e Blaise Pascal, CNRS/IN2P3, F-63177 Aubi\`ere, France}
\author{J.J.~LeRose}
\affiliation{Thomas Jefferson National Accelerator Facility, Newport News, Virginia 23606, USA}
\author{R.~Lindgren}
\affiliation{University of Virginia, Charlottesville, Virginia 22904, USA}
\author{N.~Liyanage}
\affiliation{University of Virginia, Charlottesville, Virginia 22904, USA}
\author{H.-J.~Lu}
\affiliation{Department of Modern Physics, University of Science and Technology of China, Hefei 230026, China}
\author{D.J.~Margaziotis}
\affiliation{California State University, Los Angeles, Los Angeles, California 90032, USA}
\author{Z.-E.~Meziani}
\affiliation{Temple University, Philadelphia, Pennsylvania 19122, USA}
\author{K.~McCormick}
\affiliation{Rutgers, The State University of New Jersey, Piscataway, New Jersey 08854, USA}
\author{R.~Michaels}
\affiliation{Thomas Jefferson National Accelerator Facility, Newport News, Virginia 23606, USA}
\author{B.~Michel}
\affiliation{LPC Clermont-Ferrand, Universit\'e Blaise Pascal, CNRS/IN2P3, F-63177 Aubi\`ere, France}
\author{B.~Moffit}
\affiliation{College of William and Mary, Williamsburg, Virginia 23187, USA}
\author{P.~Monaghan}
\affiliation{Massachusetts Institute of Technology,Cambridge, Massachusetts 02139, USA}
\author{S.~Nanda}
\affiliation{Thomas Jefferson National Accelerator Facility, Newport News, Virginia 23606, USA}
\author{V.~Nelyubin}
\affiliation{University of Virginia, Charlottesville, Virginia 22904, USA}
\author{M.~Potokar}
\affiliation{Institut Jozef Stefan, University of Ljubljana, Ljubljana, Slovenia}
\author{Y.~Qiang}
\affiliation{Massachusetts Institute of Technology,Cambridge, Massachusetts 02139, USA}
\author{R.D.~Ransome}
\affiliation{Rutgers, The State University of New Jersey, Piscataway, New Jersey 08854, USA}
\author{J.-S.~R\'eal}
\affiliation{LPSC, Universit\'e Joseph Fourier, CNRS/IN2P3, INPG, F-38026 Grenoble, France}
\author{B.~Reitz}
\affiliation{Thomas Jefferson National Accelerator Facility, Newport News, Virginia 23606, USA}
\author{Y.~Roblin}
\affiliation{Thomas Jefferson National Accelerator Facility, Newport News, Virginia 23606, USA}
\author{J.~Roche}
\affiliation{Thomas Jefferson National Accelerator Facility, Newport News, Virginia 23606, USA}
\author{F.~Sabati\'e}
\affiliation{CEA Saclay, DAPNIA/SPhN, F-91191 Gif-sur-Yvette, France}
\author{A.~Saha}
\affiliation{Thomas Jefferson National Accelerator Facility, Newport News, Virginia 23606, USA}
\author{S.~Sirca}
\affiliation{Institut Jozef Stefan, University of Ljubljana, Ljubljana, Slovenia}
\author{K.~Slifer}
\affiliation{University of Virginia, Charlottesville, Virginia 22904, USA}
\author{P.~Solvignon}
\affiliation{Temple University, Philadelphia, Pennsylvania 19122, USA}
\author{R.~Subedi}
\affiliation{Kent State University, Kent, Ohio 44242, USA}
\author{V.~Sulkosky}
\affiliation{College of William and Mary, Williamsburg, Virginia 23187, USA}
\author{P.E.~Ulmer}
\affiliation{Old Dominion University, Norfolk, Virginia 23508, USA}
\author{E.~Voutier}
\affiliation{LPSC, Universit\'e Joseph Fourier, CNRS/IN2P3, INPG, F-38026 Grenoble, France}
\author{K.~Wang}
\affiliation{University of Virginia, Charlottesville, Virginia 22904, USA}
\author{L.B.~Weinstein}
\affiliation{Old Dominion University, Norfolk, Virginia 23508, USA}
\author{B.~Wojtsekhowski}
\affiliation{Thomas Jefferson National Accelerator Facility, Newport News, Virginia 23606, USA}
\author{X.~Zheng}
\affiliation{Argonne National Laboratory, Argonne, Illinois, 60439, USA}
\author{L.~Zhu}
\affiliation{University of Illinois, Urbana, Illinois 61801, USA}
\collaboration{The Jefferson Lab Hall A Collaboration}

\makeatletter
\global\@specialpagefalse
\def\@oddhead{ }
\let\@evenhead\@oddhead
\def\@oddfoot{\reset@font\rm\hfill \thepage\hfill
} \let\@evenfoot\@oddfoot
\makeatother

\begin{abstract}

The present experiment exploits the interference between the Deeply Virtual Compton Scattering (DVCS) 
and the Bethe-Heitler processes to extract the imaginary part of DVCS amplitudes on the neutron and 
on the deuteron from the helicity-dependent D$({\vec e},e'\gamma)X$ cross section measured at $Q^2$=1.9~GeV$^2$ and $x_B$=0.36. 
We extract a linear combination of generalized parton distributions (GPDs) particularly 
sensitive to $E_q$, the least constrained GPD. A model dependent constraint on the contribution of the 
up and down quarks to the nucleon spin is deduced.

\end{abstract}


\maketitle

%
%
%

Understanding the structure of the nucleon in terms of quarks and gluons is a central project of modern 
hadronic physics. In the non-perturbative regime relevant to nuclear scales, Quantum Chromodynamics 
(QCD), the theory describing the elementary dynamics of the nucleon, is not yet solvable and remains 
rather mysterious. The electromagnetic probe provides an outstanding tool to study the nucleon 
structure. In this letter, we present the first study of the $({\vec e},e'\gamma)$ reaction on neutrons 
off a deuterium target.

Elastic electron scattering revealed the non-pointlike nature of the nucleon~\cite{Hof55}. Deviations 
from the Mott cross section define the electromagnetic form factors which describe the spatial 
distribution of charge and current inside the nucleon, as functions of the invariant momentum transfer 
squared ($Q^2$), i.e. the resolution of the probe. Deep Inelastic Scattering (DIS) revealed the partons 
inside the nucleon~\cite{Bre69}. The DIS cross section can be expressed in terms of the probability to 
find a quark with fraction $x_B$ of the nucleon longitudinal momentum. This motivated extensive 
measurements of the momentum distribution of quarks and gluons in nucleons, i.e. the parton 
distributions. Polarized DIS with longitudinally polarized beams and targets measures the probability 
to find a parton of given momentum with spin aligned or anti-aligned with the proton spin. These 
experiments yielded the unexpected result that the quarks carried about only 30\% of the total spin of 
the nucleon~\cite{Ash89} and questioned the role of gluons in this puzzle. Experimental results to date 
suggest that the gluon polarization does not contribute significantly to the nucleon 
spin~\cite{{Adl04},{Age06}}. A natural candidate to solve this problem is the orbital angular momentum 
of quarks and gluons.

\begin{figure}[t]
\begin{center}
\includegraphics[width=0.55\linewidth]{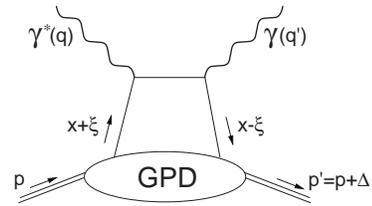}
\caption{Lowest order (QCD) amplitude for the virtual Compton process. The momentum four-vectors of the 
incident and scattered photon are $q$ and $q'$, respectively.  The momentum four-vectors of the initial 
and final proton are $p$ and $p'$, with $\Delta$=$(p'-p)$=$(q-q')$. The DIS scaling variable is 
$x_{\rm B}$=$Q^2/(2 p \cdot q)$ and the DVCS scaling variable is $\xi$=$x_{\rm B}/(2-x_{\rm B})$. In 
light cone coordinates defined by $P$=$(p+p')/2$, the initial and final momentum of the photons are 
$-2\xi$ and $0$, respectively.}
\label{fig:1}
\end{center}
\end{figure}

The Generalized Parton Distribution (GPD) framework provides a new formalism to unravel the nucleon 
structure that unifies form factors, parton distributions, and the angular momentum of 
partons~\cite{{Mue94},{Rad97},{Ji97}}. Quark GPDs are four universal functions $H_q$, $E_q$, 
${\widetilde H}_q$, and ${\widetilde E}_q$, defined by the nucleon helicity-conserving and 
helicity-flip matrix elements of the vector and axial-vector currents for quark flavor $q$. 
The GPDs correspond to the amplitude for 
removing a parton of momentum fraction $x+\xi$ and restoring it with momentum fraction $x-\xi$ 
(Fig.~\ref{fig:1}). In this process, the nucleon receives an invariant momentum transfer $t$=$\Delta^2$. 
In impact parameter space, the transverse momentum transfer $\Delta_\perp$ is Fourier conjugate to the 
transverse position of the parton leading to a femto-tomography~\cite{{Bur00},{Ral02},{Die02},{Bel02}} 
of the nucleon: GPDs represent distributions, in the transverse plane, of partons carrying longitudinal 
momentum $x$. The correlation of the position and momentum of quarks can exactly be combined into an 
orbital momentum~\cite{Bur05}. This is explicit in Ji's sum rule~\cite{Ji97} 
\begin{eqnarray}
 & J_q & = \frac{1}{2} \Delta \Sigma_q + L_q \label{Jisumrule} \\
 & = & \frac{1}{2} \int_{-1}^{+1} dx \, x \left[ H_q(x,\xi,t=0) + E_q(x, \xi, t=0) \right] \nonumber
\end{eqnarray}
which is independent of the $\xi$-value. Polarized DIS measures the spin part $\Delta \Sigma_q$, and 
unpolarized DIS determines the momentum sum rule and forward limit of $H_q$ 
\begin{equation}
M_q = \int_{-1}^{+1} dx \, x q(x) = \int_{-1}^{+1} dx \, x H_q(x,\xi=0,t=0) 
\end{equation}
meaning that the contribution of $H_q$ to Eq.~\ref{Jisumrule} is known.
Constraints on $E_q$ will allow access to the quark orbital momentum in the nucleon.

Deeply virtual Compton scattering (DVCS) is the simplest reaction to access GPDs (Fig.~\ref{fig:1}). In 
the Bjorken limit, similar to DIS, where $-t \ll Q^2$ and $Q^2$ is much larger than the quark 
confinement scale, the factorization theorem separates the reaction amplitude into the convolution of 
a known perturbative $\gamma^{\ast} q \to \gamma q$ kernel with an unknown soft matrix element 
describing the nucleon structure (GPDs)~\cite{{Ji98},{Col99}}. The Bethe-Heitler (BH) process, where the 
real photon is emitted by either the incoming or scattered electrons, serves as a reference amplitude 
that interferes with the Compton amplitude. The difference between polarized cross sections for 
opposite beam longitudinal polarization isolates, at leading order in $1/Q$, the imaginary part of the 
interference between the BH and DVCS amplitudes~\cite{Die97}. This difference is a direct measurement of 
a linear combination of GPDs~\cite{BeM02} dominated by the contribution of $E_q$ in the neutron case.

The first evidence for DVCS was reported in beam-helicity asymmetries at HERMES~\cite{Air01}, and 
CLAS~\cite{Ste01}, and in unpolarized cross sections at HERA~\cite{{Adl01},{Che03},{Akt05}}, and more
recently in measurements of the target spin asymmetry~\cite{Che06} and beam charge 
asymmetry~\cite{Air07}. A dedicated H$({\vec e},e'\gamma)p$ experiment in Jefferson Laboratory (JLab) 
Hall A shows evidence for factorization at $Q^2$ as low as 2.0 GeV$^2$~\cite{Mun06}. The E03-106 
experiment~\cite{E03-106} reported here is an exploratory experiment investigating for the first time the 
DVCS reaction off the neutron.

%
%
%

The experimental data were acquired in JLab Hall A, consecutively to the H$({\vec e},e'\gamma)p$ 
experiment. A 5.75~GeV/$c$ longitudinally polarized electron beam impinged on a 15~cm liquid D$_2$ cell 
serving as quasi-free neutron target. Scattered electrons of 2.94~GeV/$c$ were detected at 
19.3$^{\circ}$ in the left High Resolution Spectrometer (HRS-L)~\cite{Alc04} selecting kinematics at 
$Q^2$=1.9~GeV$^2$ and $x_B$=0.36. DVCS photons were detected in a PbF$_2$ electromagnetic calorimeter 
organized in an 11$\times$12 array of 3$\times$3$\times$18.6~cm$^3$ crystals centered around the 
direction of the virtual photon at $-18.3^{\circ}$. The calorimeter front face was 110~cm from the 
target center covering a $t$ acceptance $-0.5$~GeV$^2<t$. Typical beam intensities of 4~$\mu$A yielded a 
4$\times$10$^{37}$~cm$^{-2} \cdot$s$^{-1}$/nucleon luminosity with 76\% polarized electrons. Three 
independent reactions are used to calibrate and monitor the calorimeter: 
H$(e,e'_{\mathrm{Calo.}}p_{\mathrm{HRS}})$, D$(e,e'_{\mathrm{Calo.}}\pi^-_{\mathrm{HRS}})pp$, and 
H,D$(e,e'_{\mathrm{HRS}}\pi^0_{\mathrm{Calo.}})X$~\cite{Maz06}. Recoil particle detection in the HRS 
provides tagged electrons in the calorimeter allowing for the independent calibration of each block. 
The mass and width of the $\pi^0$ peak reconstructed from the invariant mass of $\gamma \gamma$ events 
in the calorimeter provide independent tests of the previous calibrations. $\pi^-_{\mathrm{HRS}}$ and 
$\pi^0_{\mathrm{Calo.}}$ data have been taken simultaneously with DVCS data, ensuring a continuous 
monitoring of the calibration and the resolution of the calorimeter. A 1\% uncertainty on the 
calorimeter calibration was estimated from the differences between $\pi^-$ and $\pi^0$ calibrations. 
The final state of the D$({\vec e},e'\gamma)X$ reaction was selected via the squared missing mass 
$M_X^2$=$(q+p-q')^2$ reconstructed from the virtual and real photons. 
 
%
%
%

\begin{figure}[t]
\begin{center}
\includegraphics[width=0.95\linewidth]{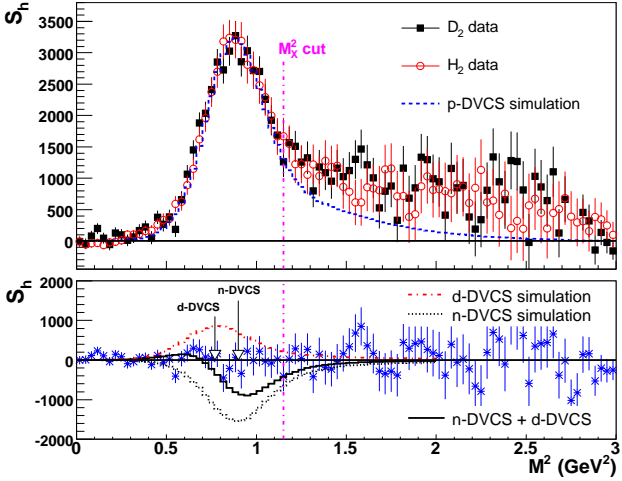}
\caption{(top) Helicity signal (Eq.~\ref{sh}) for D$(e,e'\gamma)X$ and H$(e,e'\gamma)X$ events; H$_2$ data 
are folded with a momentum distribution of the proton in deuterium, and scaled to the D$_2$ data 
luminosity; the simulation curve is for the Fermi broadened H$(e,e'\gamma)p$ reaction. (bottom) 
Residual helicity signal after H$_2$ subtraction; the arrows indicate the $M_X^2$ average position of 
n-DVCS and d-DVCS events for $<t>$=$-0.3$~GeV$^2$; the simulation curves, integrated over the complete experimental
acceptance and obtained for the arbitrary values 
$\Im \text{m} \left[{\mathcal C^I_n}\right]^{exp}$=$-\Im \text{m} \left[{\mathcal C^I_d}\right]^{exp}$=$-1$ 
(Eq.~\ref{eq:dSigmatot}), illustrate the sensitivity of the data to the neutron and deuteron signals.}

\label{asy1}
\end{center}
\end{figure}

The three-momentum transfer $|{\vec \Delta}|$ to the target varies within 0.4-0.8 GeV/c in our acceptance. 
In this range, the impulse approximation (IA) is expected to accurately describe the inclusive yield. 
Within the IA, the cross section for electroproduction of photons on a deuterium target may be decomposed 
into elastic (d-DVCS) and quasi-elastic (p-DVCS and n-DVCS) contributions following
\begin{equation}
\mathrm{D}({\vec e},e'\gamma) X = d({\vec e},e'\gamma)d + n({\vec e},e'\gamma)n + p({\vec e},e'\gamma)p + \ldots
\label{ImAp}
\end{equation}
where meson production channels are also contributing as background. Cross sections are obtained 
from D$({\vec e},e'\gamma)X$ events after subtraction of the proton quasi-elastic contribution deduced 
from measurements on a liquid H$_2$ target: the Fermi motion of bound protons is statistically added to 
the squared missing mass $M^2_X \vert_0$ of free proton data following 
$M_X^2$=$M^2_X \vert_0-2 {\vec p_i} \cdot ({\vec q}-{\vec {q'}})$ where ${\vec p_i}$ is the initial 
proton momentum in the deuteron from~\cite{Lac80}; this leads to a 3\% relative increase of the $M^2_X$ 
spectrum resolution. 

The helicity signal ($S_h$) is defined according to 
\begin{equation}
S_h = \int_{0}^{\pi} (N^{+} - N^{-}) \, d^5\Phi - \int_{\pi}^{2\pi} (N^{+} - N^{-}) \, d^5\Phi
\label{sh}
\end{equation}
where $d^5\Phi=dQ^2 dx_{\rm B} dt d\phi_e d\phi_{\gamma\gamma}$ is the detection hypervolume; the 
integration boundaries in Eq.~\ref{sh} define the limits in the azimuthal angle $\phi_{\gamma\gamma}$~\cite{Bac04}; 
$N^{\pm}$ are the number of counts for $\pm$ beam helicity, corrected for random 
coincidences, and integrated over a particular bin in $M_X^2$. The helicity signal for D$_2$ and H$_2$ 
targets from $({\vec e},e'\gamma)$ coincident detection is displayed in Fig.~\ref{asy1} (top) as a 
function of the squared missing mass. For our purposes, $M^2_X$ is calculated with a target 
corresponding to a nucleon at rest, leading to the kinematic $\Delta M_X^2 \simeq t/2$ separation 
between deuteron elastic and nucleon quasi-elastic contributions. Pion production channels 
($e A \rightarrow e A \gamma \pi$, $e A \rightarrow e A \pi^0 \pi \ldots$) are strongly suppressed by 
the kinematical constraint $M_X^2 < (M+m_{\pi})^2$=$M_X^2|_{\rm cut}$. Their contribution to the 
helicity signal of p-DVCS, induced via resolution effects below $M^2_X|_{\rm cut}$, was found to be 
negligible on the proton as illustrated by the comparison between H$_2$ data and scaled simulations 
(Fig.~\ref{asy1} top). Figure~\ref{asy1} (bottom) shows the subtraction (D$-$H data) of the two spectra 
of Fig.~\ref{asy1} (top). The residual helicity signal for $M_X^2 < M_X^2|_{\rm cut}$ is compatible 
with zero. It corresponds to the sum of the coherent d-DVCS and incoherent n-DVCS processes (Eq.~\ref{ImAp}). 
Asymmetric decays of $\pi^0$ (in $e A \rightarrow e A \pi^0$), where only one photon is detected in the 
calorimeter, mimic DVCS events. The contamination due to this background was treated as a systematic 
error estimated from the number of detected $\pi^0$ events, corresponding to primarily symmetric 
decays~\cite{Mun06}. 

%
%
%

The H$_2$ results~\cite{Mun06} show that the handbag mechanism (Fig.~\ref{fig:1}) dominates the p-DVCS 
helicity-dependent cross section difference at our kinematics. As a consequence, only twist-2 
contributions are considered in this analysis. The $exp$ superscript in Eq.~\ref{eq:dSigmatot} reflects 
this restriction. In the impulse approximation, we write the experimental helicity-dependent 
cross-section difference as the sum of the (incoherent) neutron and the (coherent) deuteron 
contributions, within the formalism of Refs.~\cite{{BeM02},{Kir03}}
\begin{eqnarray}
& & {d^5\Sigma_{{\mathrm D}-{\mathrm H}} \over d^5\Phi} = {1\over 2} \left[ {d^5\sigma^+\over d^5\Phi} - 
{d^5\sigma^-\over d^5\Phi}\right] \label{eq:dSigmatot} \\
 & = & \bigg( \Gamma_{d}^{\Im} \, 
\Im \text{m} \left[{\mathcal C^I_d}\right]^{exp} + \Gamma_{n}^{\Im} \, \Im \text{m} \left[{\mathcal C^I_n}\right]^{exp} 
\bigg) \, \sin(\phi_{\gamma\gamma}) \nonumber \  .
\end{eqnarray}
$\Gamma^{\Im}_{n,d}$ are kinematical factors with a $\phi_{\gamma\gamma}$ dependence that arises from 
the electron propagators of the BH amplitude; in the $t$ range of interest, the averaged 
$\Gamma^{\Im}_n/\Gamma^{\Im}_d$ ratio vary from 0.4 to 0.9 with increasing $|t|$. 
$\Im \text{m} \left[{\mathcal C^I_n}\right]$ depends on the interference of the BH amplitude with the 
set ${\mathcal F}=\{{\mathcal H,\,\mathcal E,\,\widetilde{\mathcal H}}\}$ of twist-2 Compton form 
factors (CFFs):
\begin{equation}
{[{\mathcal C^I_n}]}^{exp} \simeq {[{\mathcal C^I_n}]} = F_1{\mathcal H} + \xi (F_1+F_2) \widetilde{\mathcal H} - 
{t\over 4 M^2} F_2{\mathcal E} 
\label{eq:gpds1}
\end{equation} 
where $F_1$($F_2$) is the Dirac(Pauli) form factor entering into the BH amplitude. Similarly, 
$\Im \text{m} \left[{\mathcal C^I_d}\right]$ depends on the different set of spin-1 CFFs of the 
deuteron~\cite{Kir03}. The imaginary part of twist-2 CFFs is determined by the $x=\pm\xi$ points of the 
GPDs, with for example:
\begin{equation}
\Im \text{m} \left[{\mathcal E}\right] = \pi \sum_{q} e_q^2 \left( E_q(\xi,\xi,t) - E_q(-\xi,\xi,t)\right) \ .
\label{eq:CFF}
\end{equation}
where $e_q$ is the quark charge in units of the elementary charge.
While Eq.~\ref{eq:gpds1} for a proton is dominated by ${\mathcal H}$ and $\widetilde{\mathcal H}$, it  
becomes essentially sensitive to $\mathcal E$ in the neutron case following the small value of $F_1$ 
and the cancellation between $u$ and $d$ polarized parton distributions in 
$\widetilde{\mathcal H}$~\cite{Goe01}. 

\begin{figure}[t]
\begin{center}
\includegraphics[width=0.95\linewidth]{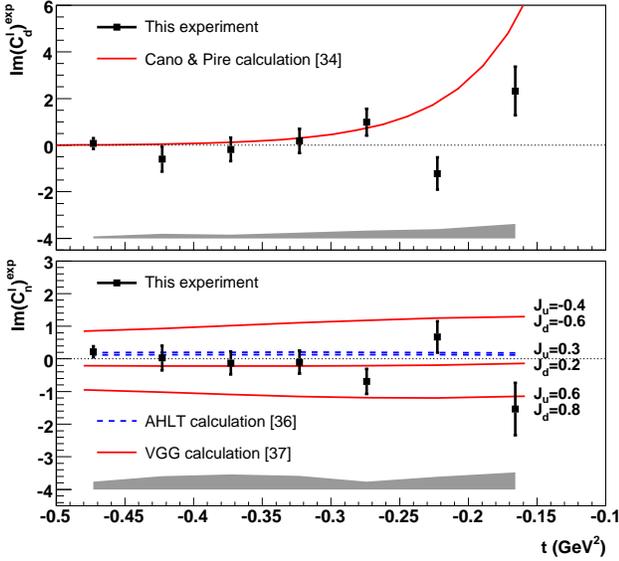}
\caption{The $t$-dependence of the extracted $\sin (\phi_{\gamma\gamma})$ moments for coherent d-DVCS 
(top panel) and incoherent n-DVCS (bottom panel). Error bars show statistical uncertainties; 
systematical uncertainties are indicated by the shaded bands.}
\label{result}
\end{center}
\end{figure}
$\Im \text{m}\left[{\mathcal C^I_n}\right]^{exp}$ and $\Im \text{m}\left[{\mathcal C^I_d}\right]^{exp}$ 
are simultaneously extracted in each $t$-bin from a global analysis involving 7$\times$12$\times$30 bins 
in $t \otimes \phi_{\gamma\gamma} \otimes M_X^2 \in [-0.5;-0.1]$~GeV$^2 \otimes [0;2\pi] \otimes 
[0.;1.15]$~GeV$^2$. A Monte Carlo simulation with the kinematic weights of Eq.~\ref{eq:dSigmatot} as a 
function of $(t,\phi_{\gamma\gamma},M_X^2)$ is fitted to the experimental distribution 
$\left[ N^+(t,\phi_{\gamma\gamma},M_X^2) - N^-(t,\phi_{\gamma\gamma},M_X^2) \right]$ obtained after the 
D$-$H subtraction. The two coefficients $\Im \text{m} \left[{\mathcal C^I_n}(t_i)\right]^{exp}$ and 
$\Im \text{m} \left[{\mathcal C^I_d}(t_i)\right]^{exp}$ are the free parameters of the fit in each bin 
$t_i$. The binning in $\phi_{\gamma\gamma}$ allows the determination of the $\sin(\phi_{\gamma\gamma})$ 
moments whereas the binning in $M_X^2$ allows the separation of the d-DVCS and n-DVCS signals. The 
simulation includes both external and real internal radiative effects. It takes also into account 
detector resolution and acceptance. Finally, virtual and soft real radiative corrections are applied with a 
global correction factor of 0.91$\pm$0.02 to the experimental yields~\cite{Van00}.
 
%
%

\begin{table}[t]
\begin{center}
\begin{tabular}{c||c|c|c}
\hline\hline
\small  $<t>$ & $\Im \text{m} \left[{\mathcal C^I_{n}}(t_i)\right]^{exp}$ & $\Im \text{m} 
\left[{\mathcal C^I_{d}}(t_i) \right]^{exp}$ & $\alpha_{nd}$ \\ 
\hline 
 -0.473 & 0.22 $\pm$ 0.17 $\pm$ 0.24 & 0.07 $\pm$ 0.23 $\pm$ 0.08  & -0.72\\ 
 -0.423 & 0.03 $\pm$ 0.38 $\pm$ 0.41 & -0.60 $\pm$ 0.54 $\pm$ 0.19 & -0.77\\
 -0.373 & -0.13 $\pm$ 0.35 $\pm$ 0.46 & 0.18 $\pm$ 0.51 $\pm$ 0.17 & -0.80\\
 -0.323 & -0.10 $\pm$ 0.35 $\pm$ 0.42 & 0.18 $\pm$ 0.52 $\pm$ 0.24 & -0.84\\
 -0.274 & -0.69 $\pm$ 0.38 $\pm$ 0.24 & 0.98 $\pm$ 0.57 $\pm$ 0.33 & -0.88\\
 -0.225 & 0.67 $\pm$ 0.48 $\pm$ 0.39 & -1.22 $\pm$ 0.69 $\pm$ 0.40 & -0.91\\
 -0.166 & -1.54 $\pm$ 0.80 $\pm$ 0.52 & 2.32 $\pm$ 1.04 $\pm$ 0.61 & -0.95\\
\hline\hline
\end{tabular}
\end{center}
\caption{Experimental values of the $\sin (\phi_{\gamma\gamma})$ moments as a function of $t$ (in GeV$^2$). The first error is statistical 
and the second is the total systematic one resulting from the quadratic sum of each contribution; $\alpha_{nd}$ is the 
correlation coefficient between the two extracted moments. }
\label{tab:results}
\end{table}
Figure~\ref{result} displays the experimental values (Tab.~\ref{tab:results}) of 
$\Im \text{m} \left[{\mathcal C^I_{n,d}}(t_i)\right]^{exp}$. At low $|t|$, the small kinematic 
separation between d-DVCS and n-DVCS leads to a strong anti-correlation between deuteron and neutron moments (Tab.~\ref{tab:results}). 
The larger statistical errors on the extraction at low $|t|$, in spite of higher absolute statistics, reflect this feature. 
The systematical errors come essentially from the $t$-dependent uncertainties on the relative calibration 
between D$_2$ and H$_2$ data, and estimates of the bound on $\pi^0$ contamination; other contributions originate 
from DVCS detectors acceptance and luminosity (3\%), beam polarization (2\%) and radiative corrections (2\%). As 
expected from Fig.~\ref{asy1}, the moments are globally compatible with zero. Experimental results are compared 
to model calculations for deuteron~\cite{{Can04},{Ber01}} and neutron~\cite{{Van99},{Ahm07}} GPDs. The deuteron 
calculations exhibit a rapid decrease of the deuteron form factors with $|t|$. The n-DVCS results are compared 
to two different models: one where the GPDs parametrization is constrained by lattice calculation of GPDs 
moments~\cite{Ahm07}, and another where $E_q$ is parametrized by the unknown contribution of valence 
quarks to the nucleon angular momentum~\cite{Goe01}. Both approaches reproduce the rather flat 
$t$-dependence of the data. Three examples of calculations corresponding to different values of the 
$u$ ($J_u$) and $d$ ($J_d$) quark contributions are shown. This comparison indicates that the present 
data provide constraints of the GPD models, particularly on $E_q$.

%
%
%

\begin{figure}
\begin{center}
\includegraphics[width=0.90\linewidth]{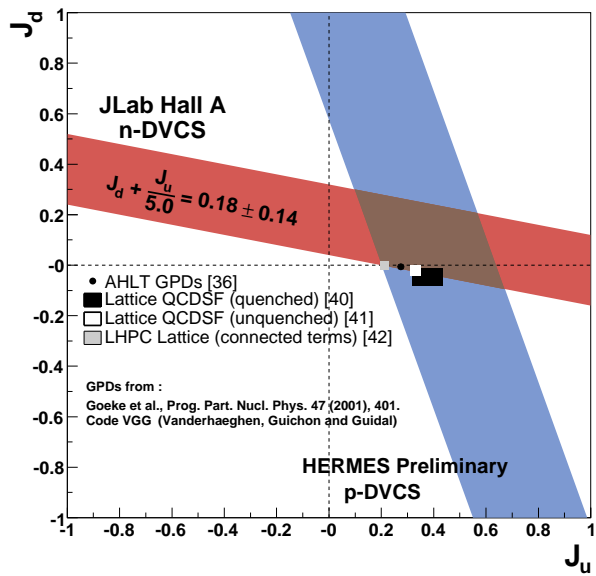}
\caption{Experimental constraint on $J_u$ and $J_d$ quark angular momenta from the present n-DVCS 
results. A similar constraint from the $\mathrm{\vec p}$-DVCS target spin asymmetry measured by 
HERMES~\cite{{Ell06},{Ye06}}, and different lattice QCD based 
calculations~\cite{{Ahm07},{Goc04},{Sch07},{Hag07}} are also shown.}
\label{JuJd}
\end{center}
\end{figure}

A correlated constraint on $J_u$ and $J_d$ can be extracted from a fit of the VGG 
model~\cite{{Goe01},{Van99}} to the neutron data (Fig.~\ref{result}), relying 
on the $\chi^2$ quantity 
\begin{equation}
\chi^2 = \sum_{i=1}^{7} \frac{\left(\Im \text{m} \left[{\mathcal C^I_n}(t_i)\right]^{exp}
-\Im \text{m} \left[{\mathcal C^I_n}(t_i)\right]^{VGG}_{J_u,J_d}\right)^{2}}{(\delta^{exp}_{stat})^2+
(\delta^{exp}_{sys})^2} \ .
\end{equation}
The condition $\chi^2 \le \chi^2_{min} + 1$ ($\chi^2_{min}$/DoF=6.6/5) defines the band 
$J_d + ( J_u / 5.0 )$~=~0.18$~\pm$~0.14 of Fig.~\ref{JuJd}. The 
model dependence of this analysis should be stressed: n-DVCS data involve GPDs at one point $x$=$\pm\xi$ and 
$t \neq 0$ while the Ji sum rule (Eq.~\ref{Jisumrule}) is an integral over $x$ extrapolated to $t$=0. A 
similar constraint obtained from HERMES preliminary $\mathrm{\vec p}$-DVCS data on a transversely 
polarized target~\cite{{Ell06},{Ye06}} is also shown together with lattice QCD based 
predictions~\cite{{Ahm07},{Goc04},{Sch07},{Hag07}}. It remains a future theoretical study to evaluate, in a model 
independent way, the constraints on $J_u$ and $J_d$ from a finite set of measurements. As expected from isospin 
symmetry, n-DVCS data have enhanced sensitivity to the $d$ quark of the proton relative to p-DVCS data. This 
complementarity is a key feature for future experimental programs investigating quark angular 
momenta.

%
%
%

In summary, this experiment provides a 
determination of the $t$-dependence of a linear combination of GPDs from the n-DVCS helicity-dependent 
cross-section difference in the range $[-0.5;-0.1]$~GeV$^2$. These data, mostly sensitive to $E_q$, were found 
to be compatible with zero. The coherent d-DVCS contribution was also extracted, the high $|t|$ behaviour 
being compatible with expectations. We provide the first experimental constraint on the parametrization of 
the GPD $E_q$ that can be expressed, within a particular model, in terms of a constraint on the quark angular 
momenta. DVCS experiments on the neutron appear as a mandatory step towards a better knowledge of the partonic 
structure of the nucleon.

%
%
%

\begin{acknowledgments}

We acknowledge essential work of the JLab accelerator staff and the Hall A technical staff. This work 
was supported in part by DOE contract DOE-AC05-06OR23177 under which the Jefferson Science Associates, 
LLC, operates the Thomas Jefferson National Accelerator Facility, the National Science Foundation, the 
French CEA and IN2P3-CNRS.

\end{acknowledgments}

%
%

\bibliography{E03-106-prl}

%
%

\end{document}